\documentclass{article}
\usepackage[a4paper, left=2cm, right=2cm, top=1.5cm, bottom=1.5cm]{geometry}
\usepackage{graphicx} 
\usepackage{amsmath}
\usepackage{physics}
\usepackage{color}

\title{\bf{Numerical simulation of autoresonant ions oscillations in an anharmonic electrostatic trap}}
\author{
{\small J. E. López\footnote{jelopdur@correo.uis.edu.co}, A. Hernández, F. F. Parada-Becerra, C. J. Paez-González, P. Tsygankov, and, E. A. Orozco\footnote{eaorozco@uis.edu.co }}\\\\
{Escuela de Física, Universidad Industrial de Santander, Bucaramanga, Colombia}
}

\date{August, 2024}
\begin{document}
\maketitle
\section*{Abstract}
This work presents the results of modelling the ion dynamics in the ART-MS (Autoresonant Trap Mass Spectrometry) device in the quasi-static approximation. This instrument utilizes an anharmonic, purely electrostatic trap for ion confinement and a radio frequency (RF) voltage source with decrementally varying frequency for selective ion extraction. The autoresonant interaction between the oscillatory motion of the ion and the RF voltage causes an increase in the amplitude of certain confined ions, allowing their selective extraction. Numerical modelling shows that the extraction of ions with a given mass occurs not only at the fundamental frequency but also at its harmonics. This effect reduces the selective properties of devices of this type because along with the main mass component for a given frequency, it is possible to enter the detector channel of ions with another mass, for which this frequency corresponds to the second or higher harmonics, even a superposition of some of these harmonics of different ions.
\\\\
\textbf{Keywords:} spectrometry, electrostatic trap, numerical modelling
\section{Introduction}
Mass spectrometry is one of the basic methods of scientific research to determine the chemical and component composition of a sample by measuring the mass-to-charge (M/Z) ratio of its ionized fragments \cite{klampfl2015direct,awad2015mass,peacock2017advances}. After total or partial ionization of the sample, the selective extraction of ions can be achieved by means of several different methods. However, the key element for their final separation is the use of some type of spatial (space-time) ion confinement in which the dynamics of these ions depend on the M/Z ratio.

In particular, electrostatic traps are currently one of the most widely used mechanisms in the manufacture and/or design of mass spectrometers, especially those based on quadrupole systems (Quadrupole Mass Spectrometers - QMS) \cite{de2007mass}. Some typical schemes employ quadrupole analyzers, in which four electrodes are used to form a quasi-harmonic electric potential well using a high-voltage radio frequency (RF) source. Time-of-flight (TOF) analyzers are another system that has become widely used. All of these spectrometers, commonly referred to as RGAs (residual gas analyzers), are usually quite large devices that are not ideally suited for in-situ chemical analysis in a variety of harsh environments \cite{dawson2013quadrupole}. This limitation arises due to two key factors: the high power consumption (20-30W) required for QMS y TOF operation and their relatively slow scan speeds. In such environments, rapid analysis with minimal power consumption are the critical requirements, so, the search for the ``ideal'' analyzer for harsh environment applications continues \cite{de2007mass,alamelu2010studies,march2010practical}. In this way, the concept of autoresonant ion confinement within an electrostatic trap was introduced by A. V. Ermakov and B. J. Hinch in 2010 as a foundation for the novel AutoResonant Trap Mass Spectrometer (ART-MS) \cite{ermakov2010electrostatic}. This innovative device employs electron ionization to produce ions, which are subsequently confined in a strongly anharmonic electrostatic potential well. Ion-selective ejection is facilitated by the application of a radio frequency (RF) voltage source with a decreasing frequency, utilizing the phenomenon of autoresonance, where the ion's natural oscillation frequency aligns with the decreasing RF frequency \cite{ermakov2010electrostatic,ermakov2011trajectory,huang2018experimental}. Compared to traditional quadrupole mass spectrometers, ART-MS offers several advantages, as identified by Brucker and Rathbone \cite{brucker2010autoresonant}. These include: (i) Faster Spectral Scans, (ii) low Power Consumption, (iii) Ultrahigh Vacuum (UHV) Compatibility, (iv) Excellent Low-Mass Ion Sensitivity, and, (v) Point Sampling.

Previous studies have shed light on the properties and phenomena underlying the mechanism. While numerical simulations confirmed ion confinement and experimental results verified the extraction process, a comprehensive investigation of the full ion dynamics is still needed to reveal the intrinsic details of both ion confinement and ion extraction using the decreasing RF source. Addressing this gap, the current work introduces preliminary efforts to numerically investigate the extraction process in ART-MS. It is expected that the results obtained will provide a valuable basis for future work on calibration and optimization of systems operating on similar principles.
\section{Theoretical formalism}
\subsection{Physical scheme}
\begin{figure}[h!]
    \centering
    \includegraphics[scale=0.56]{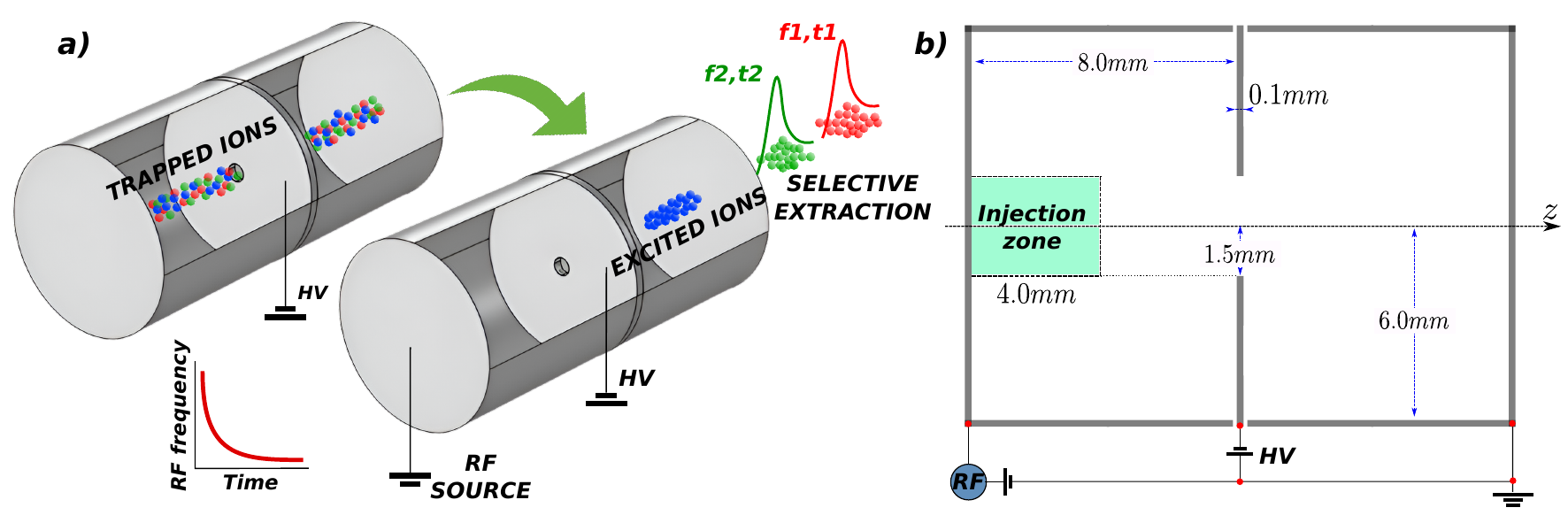}
    \caption{\small{The operational scheme of ART-MS: (a) Two grounded cylindrical electrodes and a central, flat washer electrode biased to a high negative voltage generate an anharmonic electrostatic potential well for ion confinement. When an RF source with a decreasing frequency is applied to one end of the trap, ions undergo selective excitation and extraction based on their mass-to-charge ratio. (b) A representative geometric configuration in the $y=\text{0}$ plane, showing typical dimensions: a total trap length of 16 mm and radii of 6 mm for the cylindrical electrodes. The green rectangle depicts the ion injection zone.}}
    \label{Fig_TrapScheme}
\end{figure}

At the core of ART-MS is its distinctive electrostatic ion trap, consisting of a set of electrodes. The presented minimal geometry comprises a three-electrode configuration, as illustrated in Figure \ref{Fig_TrapScheme}. A high negative DC voltage (HV) is applied to the central electrode, creating a strong confining potential well to trap the ions; which are created at one side of the trap via electron impact, with the electrons being injected with an energy range from zero to 100eV. To induce autoresonance and enable selective ion extraction, a low-amplitude RF voltage with a decreasing frequency is applied to one of the side electrodes \cite{ermakov2010electrostatic,brucker2010autoresonant}. The specific trap geometry significantly influences the performance of the instrument, commonly featuring a trap radius of 6 mm and a length of 8 mm. The central electrode, resembling a flat washer with an inner radius of 1.5 mm, typically holds a DC bias of approximately $-500$ V, while the RF voltage, about $1$ V (with decreasing frequency), is applied to one end of the trap. The interaction between the oscillating RF field and the ions' inherent oscillatory motion leads to an increase in amplitude, ultimately facilitating selective ion ejection based on the mass-to-charge ratio. Given the inverse relationship between oscillation frequency and amplitude, an extraction process can be effectively implemented using the external RF source with a decreasing frequency. This process is underpinned by the phenomenon of autoresonance,described as follows: Initially, a confined ion, characterized by a specific mass-to-charge ratio, exhibits an oscillation frequency, $\omega_o$, determined by its amplitude. Upon activating the RF source, its initial frequency, $\omega$, is set higher than $\omega_o$. The decreasing nature of the RF frequency ensures it eventually aligns with the ion's natural frequency ($\omega=\omega_o$), facilitating resonant interaction. This interaction enables the ion to gain energy, increasing its amplitude and consequently reducing its frequency to $\omega_1$. As the RF frequency further decreases to match this new frequency ($\omega=\omega_1$), the ion is excited again, leading to an increase in amplitude. This cycle of resonant interactions and energy transfer continues until the ion's amplitude reaches the trap's boundary, resulting in its extraction. The influence of the mass-to-charge ratio on oscillation frequency enables selective extraction: ions with smaller mass-to-charge ratios are extracted first, followed by ions with larger mass-to-charge ratios.
\subsection{Numerical scheme}
In this work, a simulation of the ART-MS device was conducted in three stages.  The first stage focused on the pure electrostatic trap, i.e., when the RF source voltage is turned off  (See Figure \ref{Fig_TrapScheme}). Here, the aim was to identify the electrostatic potential well and the electrostatic field responsible for ion confinement.  The second stage investigated the ion dynamics within the electrostatic trap, characterizing the behaviour of the ions and their natural oscillation frequency as a function of the mass-to-charge ratio (M/Z) and oscillation amplitude (A), with M in atomic mass units and Z the charge number.  Finally, the third stage explored the RF excitation, aiming to characterize the full ion dynamics within the device and investigate the extraction properties.
\subsubsection*{\textit{The Pure Electrostatic Trap}}
\noindent 
In the first stage, the electrostatic potential to confine the ions, $\Phi_e$, can be obtained from Laplace's equation:
\begin{equation}
    \nabla^2\Phi_e\left( \mathbf{r} \right) = 0, \label{Laplace_ES}
\end{equation}
with the Dirichlet boundary conditions: $\left.\Phi_e\right|_{\mathbf{c e}}=0$ and $\left.\Phi_e\right|_{\text {we }}=V_c$, where the subscript ``ce'' and ``cw" refer to the  \textit{cylindrical electrodes} and \textit{washer electrode}, respectively. The two cylindrical electrodes are grounded (RF source off) and the central electrode is at negative high voltage (See Figure \ref{Fig_TrapScheme}). This geometry naturally lends itself to the use of cylindrical coordinates $(r,\varphi,z)$. Due to the system's axisymmetry, Laplace's equation (\ref{Laplace_ES}) takes the form:
\begin{equation}
    \frac{1}{r}\frac{\partial\Phi_e(r,z)}{\partial r} + \frac{\partial^2\Phi_e (r,z)}{\partial r^2} + \frac{\partial^2\Phi_e (r,z)}{\partial z^2} = 0,
    \label{LaplaceEQ}
\end{equation}
In this coordinate system, a rectangular grid with uniform spacing $h_r$ and $h_z$ along the $r$ and $z$ axes is constructed, where the mesh grid-points ($r_i,z_j$) are obtained from $r_i=i h_r$ ($i=0,1,2,\cdot \cdot \cdot$) and $z_j=j h_z$ ($j=0,1,2,\cdot \cdot \cdot$). By using a finite difference scheme accurate up to second order, Equation (\ref{LaplaceEQ}) leads to:
\begin{equation}
    \Phi_e(i,j) = \frac{\left(1+\frac{1}{2i}\right)\Phi_e(i+1,j) + \left(1-\frac{1}{2i}\right)\Phi_e(i-1,j)}{2 + 2\left(\frac{h_r}{h_z}\right)^2} + \frac{\Phi_e(i,j+1) + \Phi_e(i,j-1)}{2 + 2\left(\frac{h_z}{h_r}\right)^2},
    \label{LaplaceEQFD}
\end{equation}
where $\Phi_e(i,j)$ is the electric potential calculated at the mesh gridpoint ($r_i,z_j$) \cite{sadiku2000numerical}. Equation (\ref{LaplaceEQFD}) elucidates two significant aspects:\\\\
\textbf{i)} The electrostatic potential at a specific point, $\Phi_e(i,j)$, is deduced using the values at its nearest neighbors: $\Phi_e(i+1,j)$, $\Phi_e(i-1,j)$, $\Phi_e(i,j+1)$, and $\Phi_e(i,j-1)$.\\\\
\textbf{ii)} It becomes invalid at $i=0$; i.e., at $r=0$. However, taking into account that 
\begin{equation}
    \lim_{r \to 0}\frac{1}{r}\frac{\partial \Phi_e}{\partial r} = \lim_{r \to 0}\frac{\partial^2 \Phi_e}{\partial r^2},
\end{equation}
for said limit case, Laplace's equation (\ref{LaplaceEQ}) can be written as:
\begin{equation}
    2\frac{\partial^2\Phi_e}{\partial r^2} + \frac{\partial^2\Phi_e}{\partial z^2} = 0,
\end{equation}
which can be expressed in a finite difference scheme as:
\begin{equation}
    \Phi_e(0,j)=\frac{2\Phi_e(1,j)}{1+\left(\frac{h_r}{h_z}\right)^2} + \frac{\Phi_e(0,j+1) + \Phi_e(0,j-1)}{1+\left(2\frac{h_z}{h_r}\right)^2}.
    \label{LaplaEQFDr0} 
\end{equation}
An iterative Laplace solver was developed to calculate the electrostatic potential, $\Phi_e(i,j)$, at each point of the meshgrid. This solver utilizes Equations (\ref{LaplaceEQFD}) and (\ref{LaplaEQFDr0}), iterating until the estimated potential at each point converges. Convergence is considered achieved when the change in potential between successive iterations falls below a predefined tolerance, indicating the maximal acceptable absolute difference between successive potential values \cite{sadiku2000numerical,jardin2010computational}.\\
The electric field components at the mesh-grid points, $E_r(i,j)$ and $E_z(i,j)$, are determined from the electric potential, $\mathbf{E}(r,z)=-\nabla\Phi_e(r,z)$, using a finite difference scheme. It's important to note that $E_{\varphi}(i,j)=0$ due to the axisymmetry.
\subsubsection*{\textit{Confined Ion Dynamics by the Pure Electrostatic Trap}}
\noindent The second stage focuses on the system's ability to trap ions. In this stage, the ion dynamics are numerically studied under the following assumptions: \textbf{(i)} both the electromagnetic field produced by the created ions and collisions between them and neutral atoms can be disregarded, and \textbf{(ii)}  any relativistic effects on the ion's dynamic can also be disregarded. These assumptions are based on two factors: the gas density is low ($n_{gas}\sim7.4\times10^{17}$ m$^{-3}$, due the max operating pressure, $3\times 10^{-5}$ mbar \cite{ermakov2010electrostatic}) and the maximum Lorentz factor, $\gamma_{max}$ is approximately equal to $1$, which can be calculate from the conservation energy principle: $\left(\gamma_{max}-1\right)mc^2 = U_{max}$, where $U_{max}\sim500$ eV, is the maximum potential energy. Therefore, the ion dynamic can be described by the equation:
\begin{equation}
    \frac{d^2\mathbf{r}_p}{dt^2} = \frac{q}{m}\mathbf{E}(\mathbf{r}_p), \label{Newton_Lorentz}
\end{equation}
where $\mathbf{E}(\mathbf{r}_p)$ is the electric field produced by the electrodes at the ion position, $\mathbf{r}_p$. It is worth emphasizing that ion dynamics depends on the mass-to-charge ratio, $(m/q)$. The equation (\ref{Newton_Lorentz}) can be written as the following coupled first-order differential equations:
\begin{equation}
    \frac{d\mathbf{v}}{dt} = \frac{q}{m}\mathbf{E}(\mathbf{r}_p),
    \label{Eq_motion_v}
\end{equation}
\begin{equation}
    \frac{d\mathbf{r}_p}{dt} = \mathbf{v},
    \label{Eq_motion_r}
\end{equation}
where $\mathbf{v}$ represents the ion velocity. The set of equations (\ref{Eq_motion_v})-(\ref{Eq_motion_r}) can be written in a second-order leap-frog finite difference scheme as:
\begin{equation}
    \mathbf{v}^{n+1/2} = \mathbf{v}^{n-1/2} + \frac{q}{m}\mathbf{E}(\mathbf{r}_p^{n})\Delta t,
\end{equation}
\begin{equation}
    \mathbf{r}_p^{n+1} = \mathbf{r}_p^{n} + \mathbf{v}^{n+1/2}\Delta t,
\end{equation}
where $\mathbf{r}_p^{n}=\mathbf{r}_p(t^n)$ is the ion position at time $t^n=n \Delta t$, and $\mathbf{v}^{n+1/2}=\mathbf{v}(t^{n+\frac{1}{2}})$ the ion velocity at the time $t^{n+\frac{1}{2}}=(n+\frac{1}{2}) \Delta t$; where $n=0,1,2, \cdot \cdot \cdot $ and $\Delta t$ is the time step \cite{sadiku2000numerical,decyk2023analytic,rodriguez2020implementation}. The electrostatic field value at the ion's position, $\mathbf{E}(\mathbf{r}_p^{n})$, is calculated using the bilinear interpolation method from the electric field components at the mesh-grid points, $E_r(i, j)$ and $E_z(i, j)$, described in the previous section. The azimuth angle $\varphi$ is introduced indirectly through the $(x_p, y_p)$ coordinates: $E_x=(x_p / r_p) E_r$ and $E_y=(y_p /r_p) E_r$.
\subsubsection*{\textit{Excited ions dynamics by the RF voltage source}}
\noindent In the third stage, we discuss the theoretical formalism and numerical method used to describe the ion extraction process.
The RF voltage source is connected to a cylindrical electrode (See Figure \ref{Fig_TrapScheme}), and, the RF voltage function is defined as $V_{\text{RF}}(t) = V_0\cos\theta(t)$, where $V_0$ is the RF volatge amplitude, and,
\begin{equation}
    \theta(t) = 2\pi \int_0^t f(\tau)d\tau + \theta_o.
\end{equation}
Here, $f(\tau)$ is the RF time-depend frequency function, which in the present case decreases over time, and $\theta_o$ is the initial phase.
\\
\\
Because the RF voltage source is on, there are both the electric and magnetic field components of the time-dependent electromagnetic field. However, because the condition $\ell \ll c/f $ is fulfilled; where  $\ell$ is the length of the system and $c$ is the speed of light, the quasi-electrostatic approximation can be adopted.
In the present stage, we also adopt the assumptions made in the second stage. Therefore, because the ion velocity $v \ll c$, ${F}_B<<{F}_E$, where  ${F}_B$ and ${F}_E$ are the magnetic force and the electric field on the ion, respectively \cite{zangwill2013modern,cheng1989field}. Therefore, the ion dynamic can be described by the equation:
\begin{equation}
    \frac{d^2\mathbf{r}_p}{dt^2} = \frac{q}{m}\mathbf{E}(\mathbf{r}_p,t), \label{Newton_Lorentz_2}
\end{equation}
where $\mathbf{E}(\mathbf{r}_p,t)$ is the electric field produced by the electrodes at the ion position, $\mathbf{r}_p$ in the time $t$, which is calculated from the electric
potential, $\Phi_e(\mathbf{r},t)$, as
\begin{equation}
    \mathbf{E}(\mathbf{r}_p,t) = -\nabla\Phi_e(\mathbf{r},t).
\end{equation}
The motion equation (\ref{Newton_Lorentz_2}) is solved using the numerical scheme described in the previous stages; while the electric potential, $\Phi_e(\mathbf{r},t)$, could be obtained by solving the Laplace equation, 
\begin{equation}
    \nabla^2\Phi_e(\mathbf{r},t) = 0,
\end{equation}
at each time step, with the Dirichlet boundary conditions: $\left.\Phi_e\right|_{\mathbf{c e (left)}}=V_{\text{RF}}(t)$, $\left.\Phi_e\right|_{\mathbf{c e (right)}}=0$ and $\left.\Phi_e\right|_{\text {we }}=V_c$, where the subscript ``ce(left)'', ``ce(right)'' and ``cw'' refer to the  \textit{left cylindrical electrode}, \textit{right cylindrical electrode} and \textit{washer electrode}, respectively; however it can be computationally intensive.
\\
To optimize the calculations, we take into account that the RF potential oscillates between fixed values in the range $[-V_0,V_0]$. Therefore, the Laplace equation is solved for a set of N(=100) different boundary conditions, corresponding to N(=100) $V_{RF}$ potential values. The electric potential at any time can then be obtained through a linear interpolation method from these precomputed solutions.
\section{Results and discussion}
All simulations presented in this section were conducted using a uniform spatial discretization with grid spacing, $hr=hz=0.02$ mm. To accurately capture the dynamics, the time step, $\Delta t$, was carefully selected to be 1/50th of the inverse maximum frequency present in each simulation, $\Delta t =1/50f^{sim}_{max}$.
\subsection{The Anharmonic Electrostatic Trap potential} 
\begin{figure}[h!]
    \centering
\includegraphics[scale=0.355]{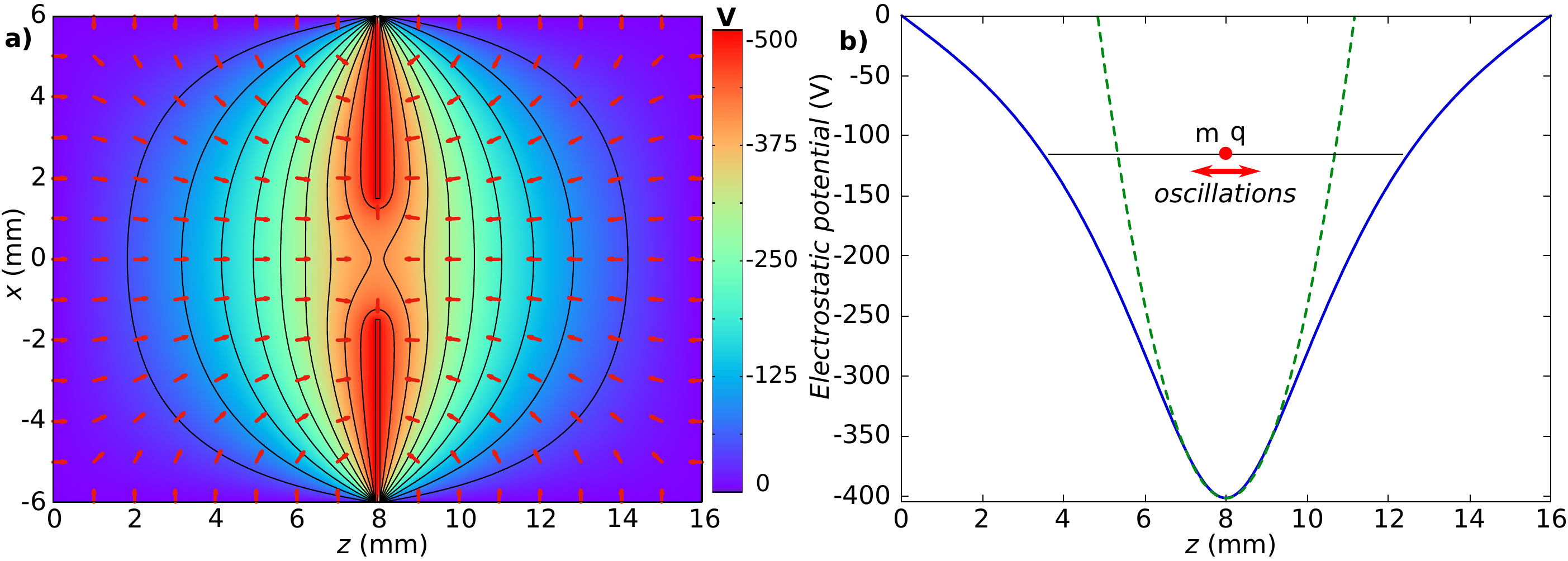}
    \caption{\small{(a) Electric potential in the $y=0$ plane, with the corresponding electrostatic field vectors $\mathbf{E}$ (red arrows). (b) Electric potential profile along the $z$-axis ($x=0$), comparing the obtained anharmonic potential well (solid blue line) with its harmonic approximation (dashed green line).}}
\label{Fig_PotentialSol}
\end{figure}

To characterize the pure electrostatic trap, Laplace's equation was solved using the following Dirichlet boundary conditions: the cylindrical electrodes are grounded and the central electrode is set to a voltage of -500 V. Figure \ref{Fig_PotentialSol}a shows the electric potential at the $y=0$ plane, where the black lines correspond to equipotential lines and the red arrows indicate the electrostatic field vectors \(\mathbf{E}\). Figure \ref{Fig_PotentialSol}b displays both the electric potential profile along the z-axis (\(x = 0\)) (the solid blue line) and the harmonic approximation (the dashed green line). The anharmonicity of the electric potential is essential for autoresonant trapping, as the ion oscillation frequency decreases with increasing oscillation amplitude, i.e., when the ions gain energy. This characteristic is crucial for selective ion extraction. In Figure \ref{Fig_PotentialSol}b the red arrow illustrate the oscillatory motion of an ion within the potential well, being $m$ and $q$ the ion's mass and charge, respectively.
\subsection{Ion dynamics in the pure electrostatic trap}
\begin{figure}[h!]
    \centering
    \includegraphics[scale=0.38]{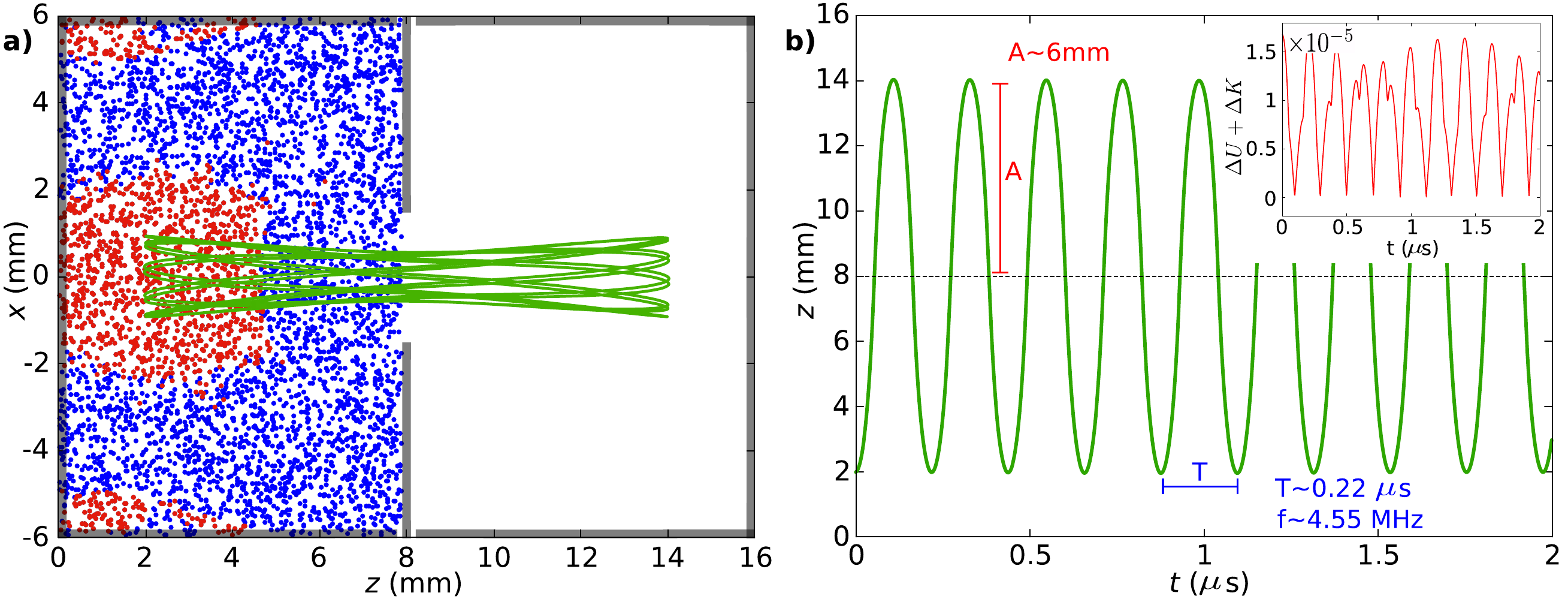}
    \caption{\small{The ion dynamics in the electrostatic trap: (a) Initial positions of the ions in the $y=0$ plane. The red dots correspond to the ions successfully confined within the electrostatic trap, while the blue dots represent the ions that impact the trap walls.The green line shows the trajectory of a trapped ion (b) Oscillatory motion of a trapped ion along the z-axis with $M/Z = 1$.}}
    \label{Fig_Trapped_ions_dynamics}
\end{figure}

To identify the injection zone that leads to stable ion oscillations, ten thousand ions were placed at rest in random positions with a uniform distribution on one side of the cavity, and the ion dynamics were simulated. Figure \ref{Fig_Trapped_ions_dynamics}a illustrates the initial positions of the ions in the $y=0$ plane. The red dots correspond to the ions successfully confined within the electrostatic trap ($\approx 16 \%$), while the blue dots represent the ions that impact the trap walls. There are two zones for the initial positions of ions where stable trajectories were obtained: (i) $r < 2$ mm and $z<5$ mm, and (ii) $r > 5$ mm and $z<5$ mm (See Figure \ref{Fig_Trapped_ions_dynamics}a). The first zone corresponds to the ion injection region shown in Figure \ref{Fig_TrapScheme}, were the atoms are ionized by electron impact. The second zone lacks practical interest, as it falls outside the performance parameters of the electrostatic autoresonant ion trap mass spectrometer under controlled conditions \cite{ermakov2010electrostatic}; therefore this zone is not considered in this study.
\\
The green line in Figure \ref{Fig_Trapped_ions_dynamics}a shows the trajectory of a representative trapped ion with an $M/Z$ ratio of 1; which exhibits oscillatory behavior, as expected for the confined ions in the trap. The oscillatory behaviour of the ion's motion along $z$-axis is shown in  Figure \ref{Fig_Trapped_ions_dynamics}b, where the amplitude of the oscillation, $A$, is approximately 6 mm, and the corresponding period, $T$, is about $0.22$ microseconds ($f\approx 4.55$ MHz).
\\
\\
To check the accuracy of the simulation results, the mechanical energy of the system was calculated as the ion describe the oscillatory motion (See inset in Figure \ref{Fig_Trapped_ions_dynamics}b). Here, $K$ and $U$ represent the Kinetic and potential energies, respectively. The maximum obtained variation of $|\Delta K + \Delta U|/U_o$ was approximately $10^{-5}$. This behavior is typical of the leap-frog scheme used in the simulation, which exhibits commendable overall stability. The small fluctuations in energy are attributed to the second-order accuracy of the numerical scheme.
\subsection{Autoresonant ion oscillation and its extraction.}
\begin{figure}[h!]
    \centering
    \includegraphics[scale=0.37]{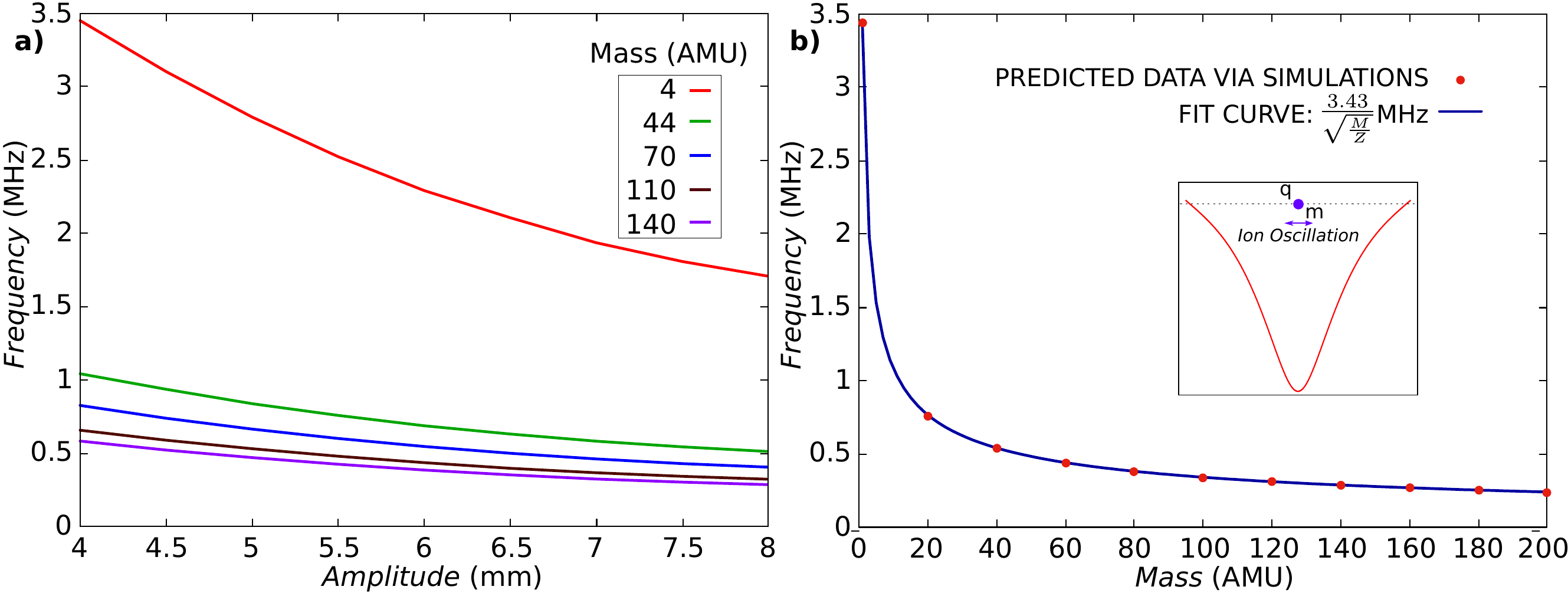}
    \caption{\small{(a) Dependence of oscillation frequency on the oscillation amplitude for various ions. (b) Approximate extraction frequency curve for an oscillation amplitude of 8 mm.}}
    \label{Fig_Fvs_A_vs_M}
\end{figure}

To analyze the dependence of oscillation frequency on the oscillation amplitude, we consider various ions with masses of 4, 44, 70, 110, and 140 Atomic Mass Units (AMU) and Z=1. The simulations were performed by varying the oscillation amplitude from 4 mm to 8 mm. (See  Figure \ref{Fig_Fvs_A_vs_M}a). The results demonstrate that the oscillation frequency decreases as the amplitude of oscillation increases for all ion masses, and the ions with larger oscillation amplitudes exhibit a lower natural oscillation frequency (NOF). This behavior, attributed to the negative anharmonicity of the potential of the electrostatic trap, is the basis of the operation of this electrostatic trap as a mass spectrometer: ions with different masses can be selectively ejected by adjusting the driving frequency.
\\
To estimate the extraction frequency as a function of mass, ion dynamics simulations were performed for ions with masses ranging from 1 to 200  AMU, in increments of 20 AMU, assuming a charge state of Z=1 and an initial amplitude of approximately 8 mm. Figure \ref{Fig_Fvs_A_vs_M}b  illustrates the relationship between the ion mass and its oscillation frequency for ions oscillating with an amplitude of approximately 8 mm. This graph provides an estimate of the extraction frequency, $f_{ext}$, for each ion. The oscillation frequency of the ions along the $z$-axis is determined solely by their mass-to-charge ratio and the amplitude of their oscillation. The frequency was calculated for ions injected along the $z$-axis at $x \approx 0$ and $y=0$, initially at rest. The predicted data points (red dots) align closely with the fitted curve (blue line), which follows the equation
    \begin{equation}
        f_{ext} (A=8 mm) \approx \frac{3.43}{\sqrt{M/Z}} \text{(MHz)}.
    \end{equation}
It is this well-predictable behavior that makes the use of such devices promising for use as a tool for mass spectrometric applications. The inset within the figure 4b illustrates a representation of the oscillatory motion of the ion within the trap. The results of the evaluations show a characteristic relationship for ART-MS systems in which, the extraction frequency is inversely proportional to the square root of the ion mass \cite{brucker2010autoresonant}. This relationship is crucial for tuning the autoresonant excitation frequency to achieve selective ion ejection, wherein by gradually tuning the RF source frequency, synchronization with specific ion extraction frequencies occurs. The analysis of these processes is important both for understanding the dynamics inside the trap, for evaluating the performance of such a device as a mass separator, and for making a theoretical evaluation of its mass resolution.
\subsection{Influence of Decay Functions on Ion Extraction in an Autoresonant Trap Mass Spectrometer}
\begin{figure}[ht]
    \centering
    \includegraphics[scale=0.37]{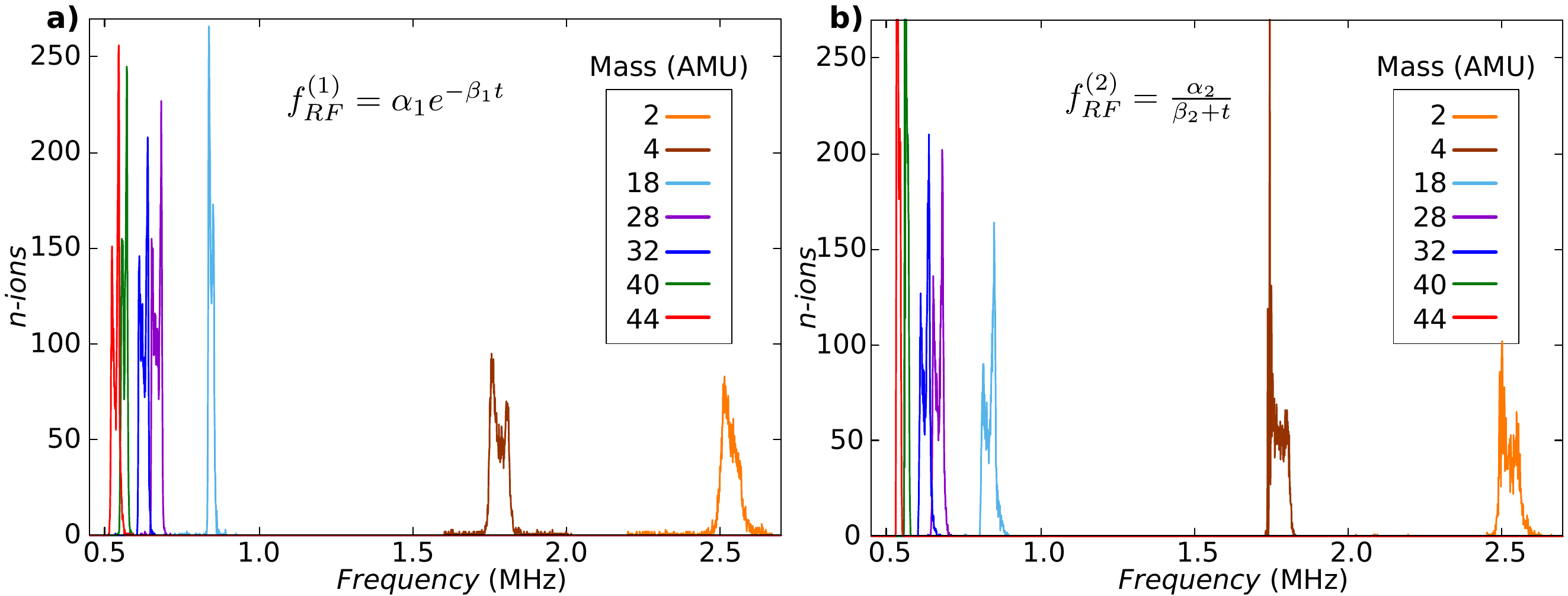}
    \caption{\small{Comparison of ion extraction results for different RF source frequency profiles: (a) Extraction using $f_{RF}^{(1)} = \alpha_1 e^{-\beta_1 t}$, and (b) Extraction using an exponential frequency decay profile $f_{RF}^{(2)} = \alpha_2/(\beta_2 + t)$. The histograms show ion number (n-ions) versus frequency (MHz) for various ion masses}}
    \label{Fig_extract_1}
\end{figure}

\begin{figure}[ht]
    \centering
    \includegraphics[scale=0.38]{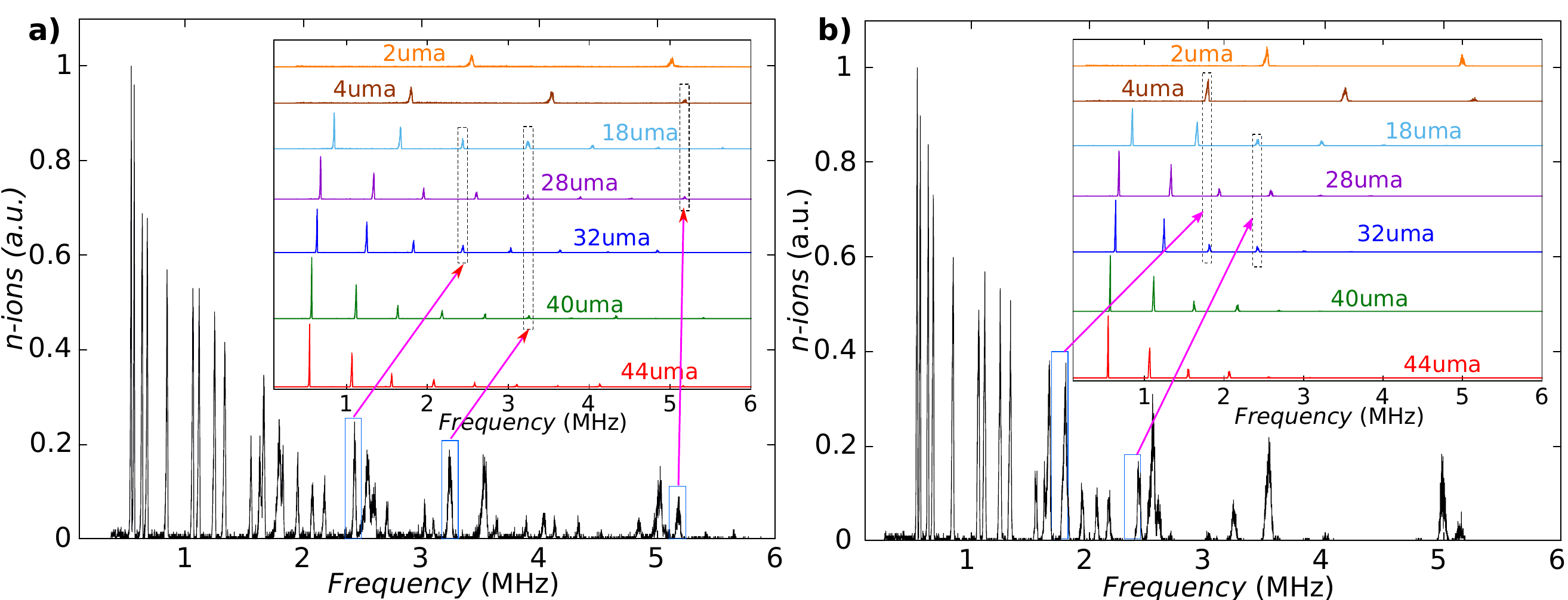}
    \caption{\small{Number of extracted ions as a function of the RF source frequency, showing multiple peaks beyond initial predictions. The figure compares ion extraction results using two different RF source frequency decay profiles. Insets illustrate extraction frequencies for different ion types, highlighting both fundamental and harmonic frequencies. Extraction reaches up to the fourth harmonic for heavier ions with \( f_{RF}^{(2)} \) and up to the eighth harmonic with \( f_{RF}^{(1)} \). Lighter ions are extracted at the fundamental and second harmonics, regardless of the frequency function. Fundamental peaks are more intense, but harmonic contributions are significant.}}
    \label{Fig_extract_2}
\end{figure}

Seven ion species with different mass-to-charge ratios, M = 2, 4, 18, 28, 32, 40 and 44 UMA, with Z=1,  were also selected for modeling the autoresonance process in a similar manner as previously selected. The corresponding extraction frequencies for an oscillation amplitude of approximately 8 mm are presented in Table \ref{Tabla_frequecies_for_extraction}. For the initial simulations, an RF source with an exponential frequency decay profile, \(f_{RF}^{(1)} = \alpha_1 e^{-\beta_1 t}\), was choosen, where \(\alpha_1=6\) MHz and \(\beta_1\approx 58.49\) s$^{-1}$ are parameters adjusted according to Emarkov's operational data. Emarkov's data depicted an ART-MS device's frequency sweep from 6 MHz to 100 kHz over approximately 70 ms. To minimize computational demands and ensure code accuracy, separate simulations for each ion species were performed over tailored intervals \((t_1, t_2)\) that encompassed sweeps around their specific extraction frequencies (Table \ref{Tabla_frequecies_for_extraction}). This method confirms successful extraction at anticipated extraction frequencies and verifies the functionality of the computational code.
\begin{table}[h!]
\begin{center}
\begin{tabular}{|cc|c|c|c|c|c|c|c|}
    \hline
    MASS       & (AMU) &   2   &   4   &  18   &  28   &  32   &  40   &  44   \\
    \hline
    $f_{ext}$      & (MHz) & 2.425 & 1.715 & 0.808 & 0.648 & 0.606 & 0.542 & 0.517 \\
    \hline
    $f_{RF}^{(1)}$ & (MHz) & 2.529 & 1.781 & 0.844 & 0.669 & 0.626 & 0.563 & 0.536 \\
    \hline
    $f_{RF}^{(2)}$ & (MHz) & 2.525 & 1.772 & 0.837 & 0.670 & 0.627 & 0.564 & 0.538 \\
    \hline
\end{tabular}
\caption{\small{Extraction frequencies for different ions according to the decay RF frequency function employed and their percentage differences with the predicted value from pure electrostatic dynamics.}}
\label{Tabla_frequecies_for_extraction}
\end{center}
\end{table}
\\
To investigate the influence of the decrease in frequency of the RF voltage source, simulations were performed using the functions $f_{RF}^{(1)} = \alpha_1 e^{-\beta_1 t}$, previously described,  and $f_{RF}^{(2)} = \alpha_2/(\beta_2 + t)$; where $\alpha_2\approx 7118.64$ and $\beta_2\approx 1.19\times 10^{-3}$ s. Figure \ref{Fig_extract_1}a shows the extraction results in separate curves for each ion species. Only ions colliding with the electrode plate at $z = 16$ mm were considered, designating this region as the extraction zone. Each graph in the Figure \ref{Fig_extract_1}a and Figure \ref{Fig_extract_1}b plots ion number ($n_{ions}$) on the y-axis vs frequency (in MHz) on the x-axis, using different colors to represent various ion masses (in AMU) with specific masses. Distinct peaks for ions with masses 2 AMU (orange) around 2.5 MHz and 4 AMU (brown) around 1.8 MHz are observed. Peaks for heavier ions, including those with masses 32, 40, and 44 AMU, are expectedly present below 1.5 MHz, indicating they are resolved at lower frequencies.
\\
The resonant excitation process is not achieved for all ions confined by the trap, approximately 20\% of each type of injected and confined ions is properly excited and extracted, indicating that this excitation and extraction mechanism occurs in approximately the same proportion for the different types of ions. Even though less intense peaks are identified for lighter ions, the width of the curve is greater than the mean width for heavier ions, making the area under the curve for each signal approximately the same. The results of the previous simulations not only corroborate the extraction mechanism proposed by Emarkov and Hinch but also validate the simulation scheme proposed in the present work.
\\
The results, presented in Figure \ref{Fig_extract_1}b and Table \ref{Tabla_frequecies_for_extraction}, associated with the frequency function $f_{RF}^{(2)}$, show no significant changes compared to those obtained with the exponential decay function, $f_{RF}^{(1)}$, although slight variations in the curve width were observed for some ion types. The central frequencies depicted in each curve align closely with those estimated from pure electrostatic trap dynamics, validating our numerical simulation approach.
\\
\\
To investigate in detail the dynamics of the ions during the entire device operation time (70 ms), the dynamics of a gas consisting of the 7 types of ions presented above were simulated, with these ions injected in equal proportion. Figure \ref{Fig_extract_2} shows two graphs comparing the number of extracted ions as a function of RF source frequency using two different decay profiles. These graphs reveal the presence of many peaks that were not expected to appear in the initial predictions. To explain this result, consider the graphs related to each ion type presented in the insets. These graphs illustrate the extraction frequencies for the different ion types, showing that the extraction occurs at the previously established and presented frequencies (See Tabla 1) and at frequencies close to multiples of these, i.e., at harmonics of the expected frequency.
\\
Extraction using the frequency decay profile \( f_{RF}^{(1)} = \alpha_1 e^{-\beta_1 t} \) is presented in Figure \ref{Fig_extract_2}a, where multiple peaks are observed across the frequency spectrum, indicating the extraction of different ion types at their fundamental and harmonic frequencies. The highest and most defined peaks correspond to the fundamental frequencies, while the harmonics are less intense but still noticeable. Lighter ions, such as those with a mass of 2 AMU, show extraction peaks at higher frequencies, whereas heavier ions, like those with a mass of 44 AMU, show peaks at lower frequencies. Ion extraction occurs up to the eighth harmonic for heavier ions when the \( f_{RF}^{(1)} \) function is considered. Figure \ref{Fig_extract_2}b illustrates that ion extraction occurs up to the fourth harmonic for simulations with the $f_{RF}^{(2)} = \alpha_2/(\beta_2 + t)$ function.
\\
The significance of harmonic frequency extraction is evident from the estimated ratios $k_1= ni(f_{ext})/ni(2f_{ext})\sim 0.6$, $k_2= ni(f_{ext})/ni(3f_{ext})\sim 0.2$, and $k_3= ni(f_{ext})/ni(4f_{ext})\sim 0.13$, where $n_i(f)$ denotes the ion intensity average at extractions frequencies $f$. Notably, the third harmonic exhibits a peak amplitude exceeding 10\% of the fundamental frequency, $f_{\text{ext}}$. A final and important observation is that some peaks are relevant due to the superposition of harmonics of different ion species, which could be mistakenly interpreted as a signal of an ion not presented in the gas. It is possible to note that for light ions, extraction occurs at their fundamental frequency and second harmonic independently of the frequency function used; and it is possible to note a not significant difference between the two frequency decay profiles. All of these observations obtained via computer simulations can help develop methodologies for calibration and avoid measurement errors.
\section{Conclusions}
A simulation of the behaviour of ions with different charge-to-mass ratios was performed for the proposed compact autoresonant trap mass spectrometers (ART-MS) \cite{ermakov2010electrostatic}. The simplest two-cavity cylindrical electrostatic trap configuration with electron injection from the end of the device was used for the analysis, where the operational parameters allow to model of the full ion dynamics via a simple electrostatic approximation.
\\
For the considered configuration of the trap the spatial trapping zones of ions with different M/Z ratios and their trajectories were visualized. The dependence of the external extraction frequency $f_{ext}$ on the mass of the extracted ion similar to previous resports is obtained \cite{ermakov2010electrostatic,ermakov2011trajectory}.
\\
Simulation results for the two investigated decay functions of external RF excitation: \( f_{RF}^{(1)} = \alpha_1 e^{-\beta_1 t} \) and $f_{RF}^{(2)} = \alpha_2/(\beta_2 + t)$, confirm the performance and potentially high mass selectivity for mass spectrometers of this type. However, the analysis showed that along with the basic harmonic $f_{ext}$, the extraction of ions of a given mass can occur at higher harmonics of this frequency. So for lighter ions extraction occurs both at the frequencies of the main and second harmonics, regardless of the frequency function used, while for heavier ions extraction extends to higher harmonics, especially when using the function $f_{RF}^{(1)}$. The ratio of extracted ions of the same mass at different harmonics can reach $k_1= ni(f_{ext})/ni(2f_{ext})\sim 0.6$, $k_2= ni(f_{ext})/ni(3f_{ext})\sim 0.2$, and $k_3= ni(f_{ext})/ni(4f_{ext})\sim 0.13$. The observed effect reduces the selectivity of the device as a mass separator and can lead to significant measurement errors. The results demonstrate the importance of considering harmonics in the design and operation of these
devices to maximize extraction efficiency and improve mass resolution. These findings have direct relevance to the optimization of ART-MS technology in analytical and research applications.
\section{Acknowledgments}
This work was carried out with the support of the Departamento Administrativo de Ciencia, Tecnolog\'ia e Innovaci\'on, Colombia, and Ministerio de Ciencia Tecnolog\'ia e Innovaci\'on, Colombia, through the announcement No 582-2019 (1102-852-69674) and the Universidad Industrial de Santander (UIS), Colombia, (Project ID: 9483-2666).

\bibliographystyle{unsrt}
\bibliography{bibliography}
\end{document}